\documentclass[mmnp]{edpsmath}
\usepackage{graphicx}
\usepackage{subcaption}
\usepackage{float}
\usepackage[many]{tcolorbox}
\definecolor{revcolor}{RGB}{0,0,0} 
\newcommand{\revised}[1]{\color{revcolor}{#1}\color{black}{}}
\begin{document}
\title{Hysteretic behavior of spatially coupled phase-oscillators}
%
\author{Eszter Feh\'er}
\address{MTA-BME Morphodynamics Research Group}
\secondaddress{Department of Mechanics, Materials and Structures, Faculty of Architecture, Budapest University of Technology and Economics}
\author{Bal\'azs Havasi-T\'oth}\address{Department of Hydraulic and Water Resources Engineering, Faculty of Civil Engineering, Budapest University of Technology and Economics}
\author{Tam\'as Kalm\'ar-Nagy}\address{Department of Fluid Mechanics, Faculty of Mechanical Engineering, Budapest University of Technology and Economics}

%
\begin{abstract}
Motivated by phenomena related to biological systems such as the synchronously flashing swarms of fireflies, we investigate a network of phase oscillators evolving under the generalized Kuramoto model with inertia. A distance-dependent, spatial coupling between the oscillators is considered. Zeroth and first order kernel functions with finite kernel radii were chosen to investigate the effect of local interactions. The hysteretic dynamics of the synchronization depending on the coupling parameter was analyzed for different kernel radii. Numerical investigations demonstrate that (1) locally locked clusters develop for small coupling strength values, (2) the hysteretic behavior vanishes for small kernel radii, (3) the ratio of the kernel radius and the maximal distance between the oscillators characterizes the behavior of the network.
\end{abstract}
%
%
%
\maketitle

\section{Introduction}
Synchronization is a collective behavior observed in many fields. It is a result of the interaction between oscillators capable of adjusting their rhythms/natural frequencies. To model synchronization of coupled phase oscillators the Kuramoto model proposed in 1975 \cite{Kuramoto1975} is often used, for example, to investigate the collective behavior of lasers \cite{Jiang1993}, neurons \cite{Cumin2007,Niyogi2009,Maistrenko2007}, social groups \cite{Yuan2017} and even crickets \cite{Walker1969}. This model was also used to describe the interesting phenomena related to \emph{Pteroptyx malaccae}, a species of firefly capable of synchronous firing with almost no phase lag \cite{Ermentrout1991}. This can be attributed to this insect's ability to alter its flashing frequency in response to external stimulus \cite{Hanson1978}. Motivated by behavior of fireflies, the original first order Kuramoto model was later extended with an inertial term by Tanaka et al. \cite{Tanaka1997} which allows for the adaptation of the flashing frequency of one firefly. They showed that in a fully coupled system, the degree of synchrony depends on the coupling strength between the oscillators in a hysteretic manner. The critical coupling strength necessary for the system to transition from the incoherent state to the coherent state is larger than the critical coupling strength resulting in the breaking of synchrony. According to Tanaka et al., the coupling strength of a firefly depends on the ratio of the brightness of the firing and that of the environment. This can lead to the swarm dynamics to exhibit hysteretic behavior. As a result, if the brightness of the background is too large, the fireflies are unable to synchronize. However, the model suggests that by increasing the brightness of the background of a synchronously flashing swarm, they can also maintain synchrony in a much brighter environment. 

The generalized model is used to describe the synchronization of Josephson junctions \cite{Trees2005} and power-grids \revised{\cite{Salam1984,Filatrella2008,Rohden2012,Olmi2014,Odor2018,Motee2014}}. The hysteretic behavior in the model is affected by many factors. In the original Kuramoto model, the coupling between the oscillators is usually considered to be undirected: they are either connected or not, which is a valid model for some applications such as power grids. Direction of current research include considering the topology of the network \cite{sieber2011stability}, the heterogeneity of the network connections \cite{Paissan2007}, the effect of dilution \cite{Olmi2014,Tumash2018} and assortativity \cite{Peron2015}. Heterogeneity can be also considered by assuming time-delay or frequency-weighted coupling \cite{Xu2016,Wu2018}. Spatial distribution of the oscillators can also result in the delay or weakening of the signal. As it was reported in \cite{Ermentrout1984}, \emph{P. malaccae} has a 3 feet range of vision which advocates the assumption of local interactions. The distance-dependency of the coupling strength was taken into account in the original Kuramoto model by using kernel functions \cite{Breakspear2010, Cenedese2015} \revised{and it was also investigated for the second order model by \cite{Odor2018,Motee2014}}. \revised{Spatial coupling of oscillators is also examined in \cite{Kapitaniak2014} considering the one dimensional coupling of neighboring pendula. Although homogeneous coupling was considered, the spatial coupling resulted in chimera states and symmetry breaking.} 
 
 In this work, we numerically analyzed the hysteretic behavior of spatially coupled phase oscillators described by the generalized Kuramoto-model with inertia. The paper is structured as follows. In the next Section, we briefly describe the model and our approach to include spatial coupling. We describe the simulations in Section 3, which is followed by the report of our main results in Section 4. Finally, in Section 5, we summarize our observations and discuss some possible applications. 
\section{The phase-oscillator model}
\subsection{Coupled oscillators with inertia}
Also known as the damped driven pendulum model, the system of coupled oscillators with the inertial term extension has been introduced and numerically investigated by \cite{Tanaka1997}. The system with $N$ mutually coupled oscillators reads as follows:
\begin{equation} \label{eq:inertia_kuramoto}
\revised{m\ddot\theta_i+\dot\theta_i=\Omega_i+\frac{K}{N}\sum_{j=1}^N{\sin(\theta_j-\theta_i)} \quad\quad\quad i=1,..,N,}
\end{equation}
where $\theta_i(t)$ and $\Omega_i$ are the phase and natural frequency of the $i$th oscillator respectively, $K$ is the coupling strength parameter expressing how quickly an oscillator can adapt to (the resultant of the) external stimuli, and $m$ is the inertial constant.

The global synchronization of the system of oscillators can be characterized by the complex order parameter 
\begin{equation} \label{eq:sync_check}
Re^{\rm{i}\Phi}=\sum_{j=1}^N{e^{\rm{i}\theta_j}},
\end{equation}
where magnitude of the complex parameter $R$ -- hereinafter referred to as order parameter -- describes the level of synchronization, $\Phi=1/N\sum_j{\theta_j}$ is the arithmetic mean of the $\theta$'s. Using (\ref{eq:sync_check}), $R=1$ corresponds to the complete synchronization ($\Phi=\theta_j$) and $R=0$ corresponds to an incoherent state of the oscillators. 

Varying the coupling strength $K$ in the system leads to a hysteretic behavior in the model. For small $K$ values, the phases of the oscillators are incoherently distributed and $R\approx0$. Increasing $K$ leads to the appearance of phase-locked oscillators and eventually the system reaches a coherent state. However, decreasing $K$ from a completely synchronized state results in a higher level of synchronization for smaller $K$ values. 

\subsection{Coupling with spatial collocation}

The second term on the right-hand side of Eq. (\ref{eq:inertia_kuramoto}) represents the Kuramoto-synchronization term, which is often modified with the adjacency matrix $A_{ij}$ so that
\begin{equation} \label{eq:kuramoto_Aij}
\frac{K}{N}\sum_{j=1}^N{A_{ij}\sin(\theta_j-\theta_i)},
\end{equation}
where the value of $A_{ij}$ is $1$ if and only if the $i$th and $j$th oscillators are coupled, and $0$ otherwise \cite{Li2015}. In case $A_{ij} \neq 1 \, \forall \, i,j$, the system is diluted. It is known, that some diluted systems exhibit a hysteretic behavior similar to the fully coupled case. The hysteretic dynamics of randomly diluted systems has recently been investigated \cite{Olmi2014}. However, even in case of the modified model Eq. (\ref{eq:kuramoto_Aij}), former investigations neglect the effect of spatial distribution on the hysteretic phase-synchronization dynamics. 

As a generalization of the mean in Eqs. (\ref{eq:inertia_kuramoto}) and (\ref{eq:kuramoto_Aij}), we propose a spatial averaging technique for the computation of pairwise coupling strength as a function of internodal distances and local neighborhoods. Systems with spatially distributed phase-oscillators may require special treatment depending on the spatial distribution. Recently \cite{Breakspear2010} and \cite{Cenedese2015} investigated the phase-oscillator model without inertial term using wavelet-like and bell-shaped kernel functions, respectively. Both works consider the kernel as a function of the internodal distances for the weighting of the coupling strength. 

We consider a set of $N$ spatially distributed nodes on the plane with positions $\textbf{r}_{i}$ with internodal Euclidean distances $d_{ij}=\vert\textbf{r}_{ij}\vert=\vert\textbf{r}_i-\textbf{r}_j\vert$. Using the nodal positions, we define the phase assigned to each node as
\begin{equation}
\theta_i=\theta(\textbf{r}_i).
\end{equation}
In order for the pairwise coupling strength to be scaled as a function of the distances, we define a spherically symmetric kernel-function $W_{ij}=W(d_{ij},\Delta)$ with (finite or infinite) smoothing radius $\Delta$ and construct the weighted average for the Kuramoto phase-synchronizer term as
\begin{align} \label{eq:weighted_mean}
&K\sum_{j=1}^{n_i}{\sin(\theta_j-\theta_i)\hat W_{ij}}, \\
&\hat W_{ij}=\frac{W_{ij}}{\sum_{j=1}^{n_i}{W_{ij}}}, \label{eq:shepard}
\end{align}
where $n_i$ is the number of neighbors within the kernel radius $\Delta$ around the $i$th node. The $j$th oscillator is a neighbor of the $j$th oscillator if it is in the $\Delta$ neighborhood of the $i$th oscillator, i.e. if $d_{ij} \leq \Delta$.
The normalization (\ref{eq:shepard}) is also known as Shepard's correction of the weighted summation \cite{Shepard1968}. 
As a result, instead of Eq. (\ref{eq:inertia_kuramoto}) we consider the following equation
\begin{equation} \label{eq:inertia_kuramoto_kernel}
\frac{\ddot\theta_i}{m}+\dot\theta_i=\Omega_i+K\sum_{j=1}^{n_i}{\sin(\theta_j-\theta_i)\hat W_{ij}} \quad\quad\quad i=1,..,N,
\end{equation}
where $K$ is the coupling parameter. Note, that in Eq. (\ref{eq:inertia_kuramoto_kernel}), the coupling parameter of the $i$th and $j$th oscillator is weighted with the kernel function, therefore the actual coupling strength between the oscillators is varying.   

The model described by Eq. (\ref{eq:inertia_kuramoto_kernel}) is a generalization of Eq. (\ref{eq:inertia_kuramoto}). Choosing $W_{ij}=1$, in the $\Delta\rightarrow \infty$ limit we have $\sum_{j=1}^{n_i}{W_{ij}}=N$ and Eq. (\ref{eq:inertia_kuramoto_kernel}) reproduces the conventional Kuramoto model. Consequently, we keep the definition of the order parameter $R$ (Eq. (\ref{eq:sync_check})) to describe the level of synchronization of the spatially coupled system.

\section{Simulations}
We implemented the model in Nauticle, the general purpose particle-based simulation tool \cite{Toth2017}. Facilitating the implementation and application of meshless numerical methods, Nauticle provides the sufficient flexibility in building arbitrary mathematical models with free-form definition of governing equations. Being a meshless open source simulation package, it is capable to solve large system of coupled ordinary differential equations, hence, the numerical model discussed in the present work can be configured easily in terms of both the equations and the geometrical layout.
 
Although the proposed model is suitable for arbitrary spatial distribution of the oscillators, as an initial study, unless stated otherwise, we investigate $N=n^2$ coupled phase oscillators placed on a two-dimensional equidistant square-grid with grid cell size $\Delta x$ and of edge length $L = (n-1)\Delta x$. Consequently, the position vector of an oscillator is $\mathbf{r}^{jk}=((j-1)\Delta x,(k-1)\Delta x)$, where $j,k=1,...,n$. Each oscillator is assigned with a scalar index parameter $i$, such that $i=(k-1)n+j$.

Following \cite{Tanaka1997}, we chose an evenly spaced natural frequency distribution on the interval of $[-\Omega_M, \Omega_M]$ such that 
\begin{equation} \label{eq:omega}
\Omega_i=-\Omega_M+\frac{2(i-1)\Omega_M}{N-1},
\end{equation}
where $\Omega_M$ is a constant. 
We considered two types of initial conditions of Eq. (\ref{eq:inertia_kuramoto_kernel}), the uniformly diffused (IC 1) taking
\begin{equation} \label{eq:inital_condition_up}
\begin{split}
&\theta_i(t=0)=2\pi\frac{i}{N}, \\
&\dot\theta_i(t=0)=\Omega_i,
\end{split}
\end{equation}
and the perfectly synchronized (IC 2), where
\begin{equation} \label{eq:inital_condition_down}
\begin{split}
&\theta_i(t=0)=0, \\
&\dot\theta_i(t=0)=0.
\end{split}
\end{equation}


In the present work we performed the simulations with two different, finite width spatial kernel-functions with a kernel radius $\Delta$. The zeroth order kernel-function is constant in the neighborhood of the $i$th oscillator
\begin{equation} \label{eq:zeroth_order}
W^0_{ij}=
\begin{cases}
&1 \quad \rm{if} \quad d_{ij}\leq\Delta, \\
&0 \quad \rm{if} \quad d_{ij}>\Delta,
\end{cases}
\end{equation}
and the first order kernel-function depends linearly on the distance between the $i$th and $j$th oscillators
\begin{equation} \label{eq:first_order}
W^1_{ij}=
\begin{cases}
&1-\frac{d_{ij}}{\Delta}, \quad \rm{if} \quad d_{ij}\leq\Delta, \\
&0, \quad \rm{if} \quad d_{ij}>\Delta,
\end{cases}
\end{equation}
consequently it is maximal at the center and decreases toward the boundary of the neighborhood. 

During the solution we applied the classic fourth order Runge-Kutta scheme for numerical integration of the system defined by Eq. (\ref{eq:inertia_kuramoto_kernel}).
We fixed the time step size $\Delta t = 0.1$ in all simulations. On the one hand, to let the value of $R$ reach a developed state, we run all cases for $500$ in simulation time and in order to eliminate the oscillations of $R$ in time, we computed the temporal average of the order parameter $\bar{R}$ of the last $2000$ steps. Also, an extended simulation of $10000$ steps had been performed to check if the results sufficiently converged but no significant changes were observed.

\section{Results}
Since the hysteretic behavior of the fully coupled system of oscillators with inertia was studied in \cite{Tanaka1997}, it is known that the temporal development of the order parameter $R$ may strongly depend on the initial conditions. In this section we present our numerical investigation in terms of the effects of the finite width spatial covering on the synchronization of the spatially distributed oscillators. We kept $\Omega_M=5$ and the inertia $m=0.85$ constant in all simulation cases and varied the kernel radii and the spatial distribution.

As a measure of the distance-based dilution we introduce a dimensionless parameter $q$, the \emph{relative kernel radius}
\begin{equation}
     q := \frac{\Delta}{d_{ij}^{max}}
\end{equation}
expressing the ratio between the kernel radius and $d_{ij}^{max}$ the maximal Euclidean distance between two oscillators of the system. In case of the zeroth order kernel-function $W_{ij}^0$ and $q\geq 1$, i.e. $\Delta>L$ the numerical setup reproduces the fully coupled second-order model (Eq. (\ref{eq:inertia_kuramoto})). 

\subsection{The effect of spatial coupling}

In order to present the local synchronization effects, we solved the equation for IC 1 and IC 2 for three different values of $q$ and both zeroth and first order kernel-functions (Eqs. (\ref{eq:zeroth_order}) and (\ref{eq:first_order}), respectively). 

\begin{figure}[H]
    \centering
    	\begin{subfigure}[c]{0.47\textwidth}
            \includegraphics[width=\textwidth]{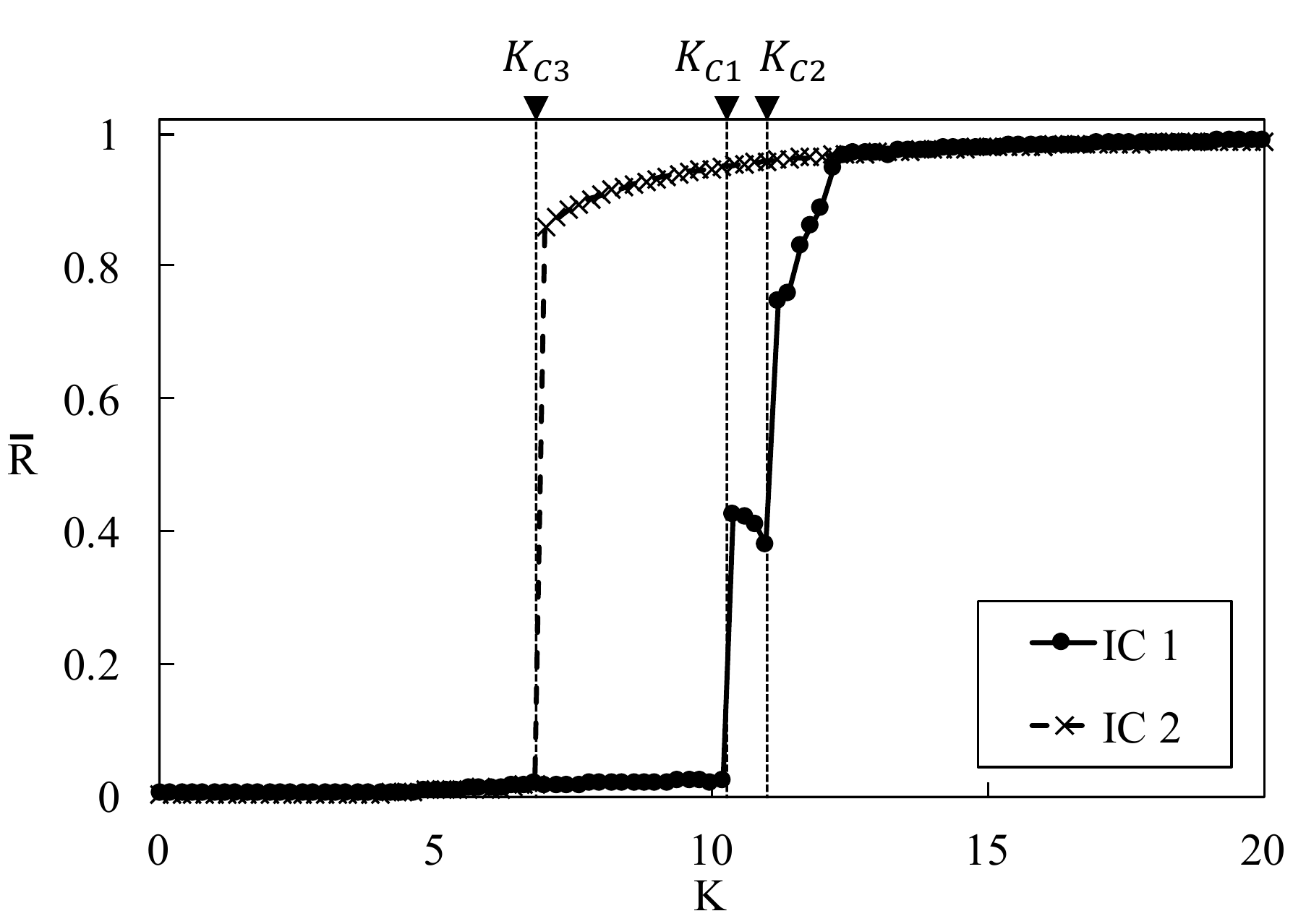}
            \caption{$\Delta = 1 \, (q\approx 0.70) $}
		    \label{}
        \end{subfigure}
        \quad
        \begin{subfigure}[c]{0.47\textwidth}
            \includegraphics[width=\textwidth]{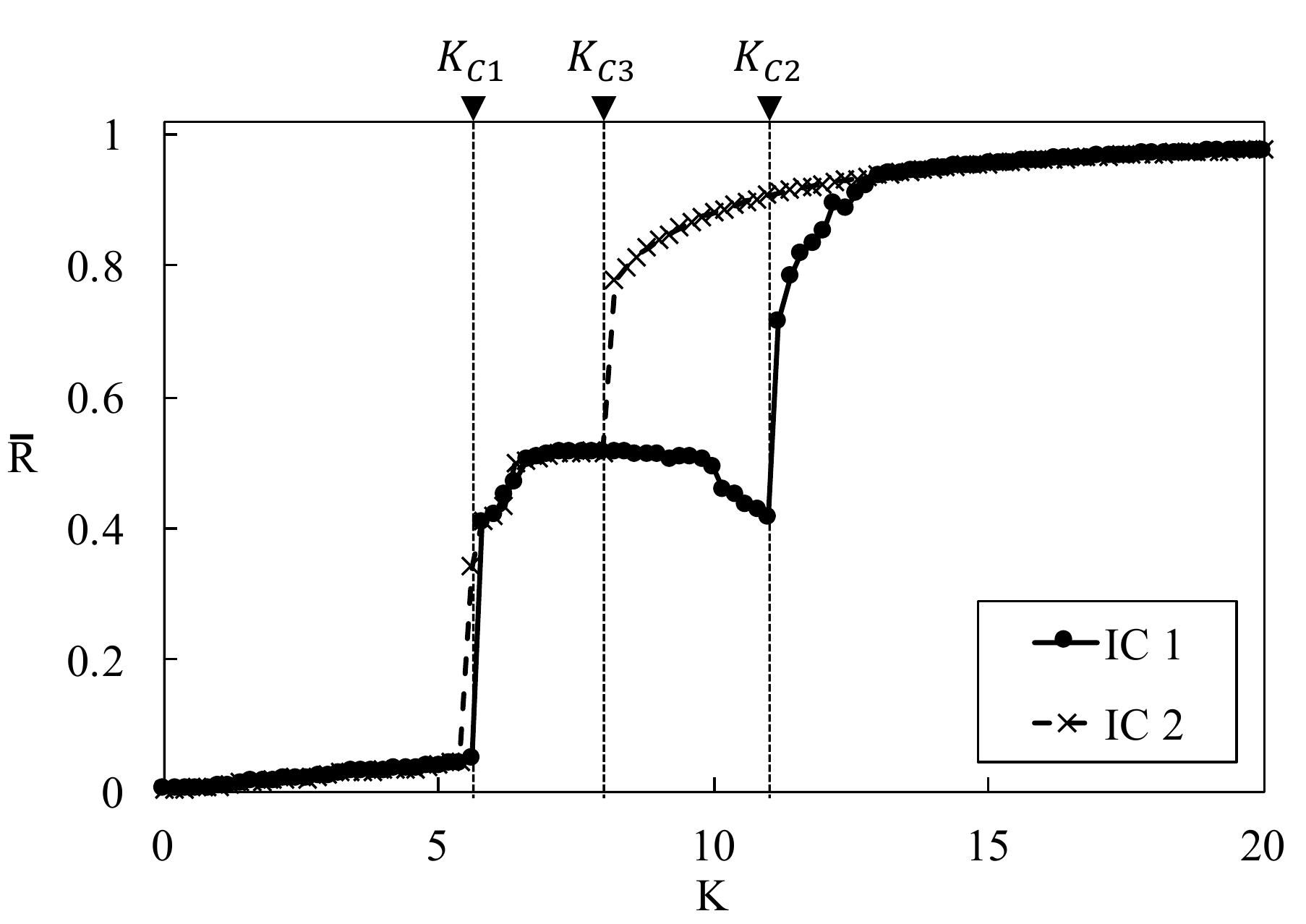}
            \caption{$\Delta = 1 \, (q\approx 0.70)$}
		    \label{}
        \end{subfigure}
            	\begin{subfigure}[c]{0.47\textwidth}
            \includegraphics[width=\textwidth]{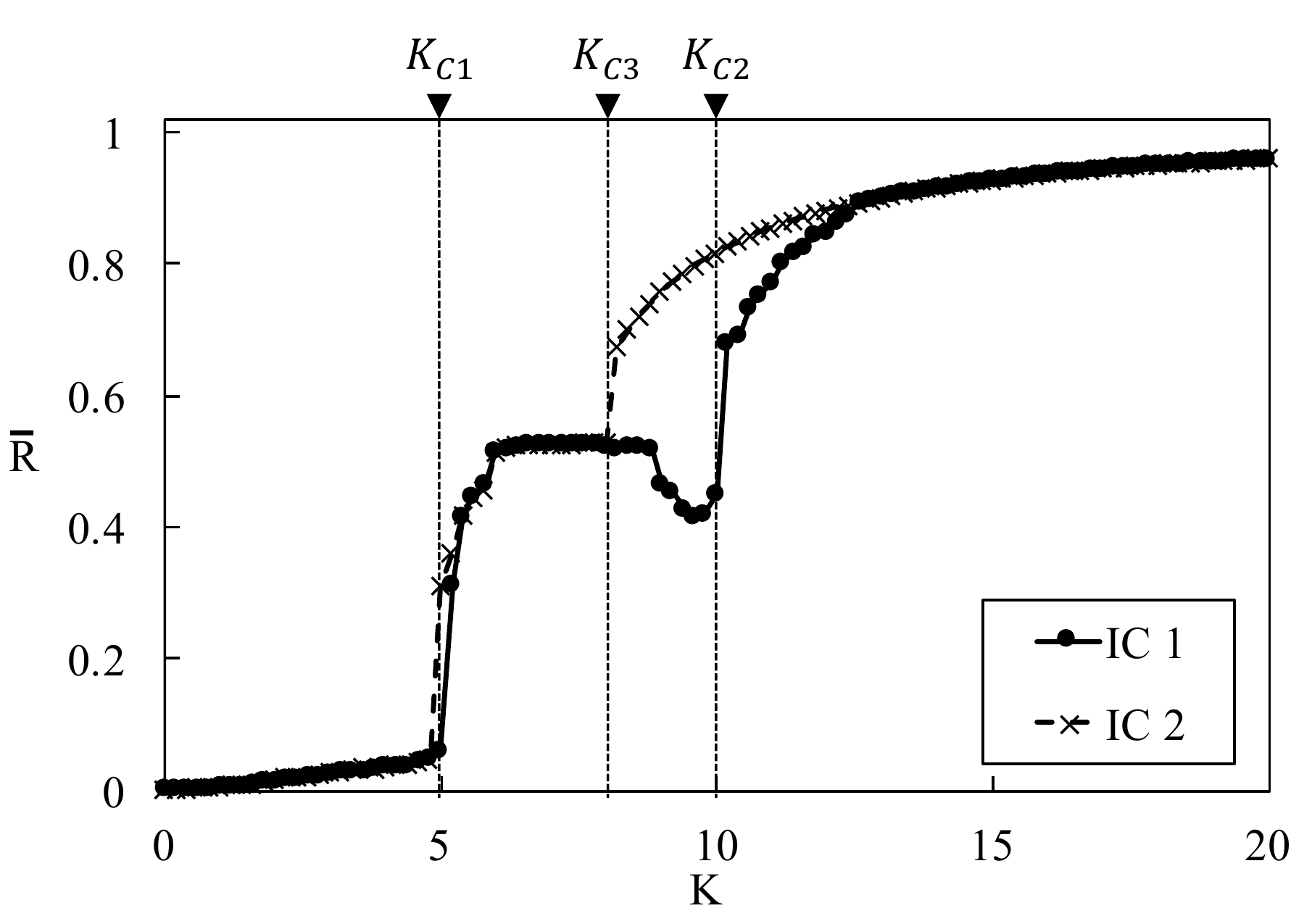}
            \caption{$\Delta = 0.6 \, (q\approx 0.42)$}
		    \label{}
        \end{subfigure}
        \quad
        \begin{subfigure}[c]{0.47\textwidth}
            \includegraphics[width=\textwidth]{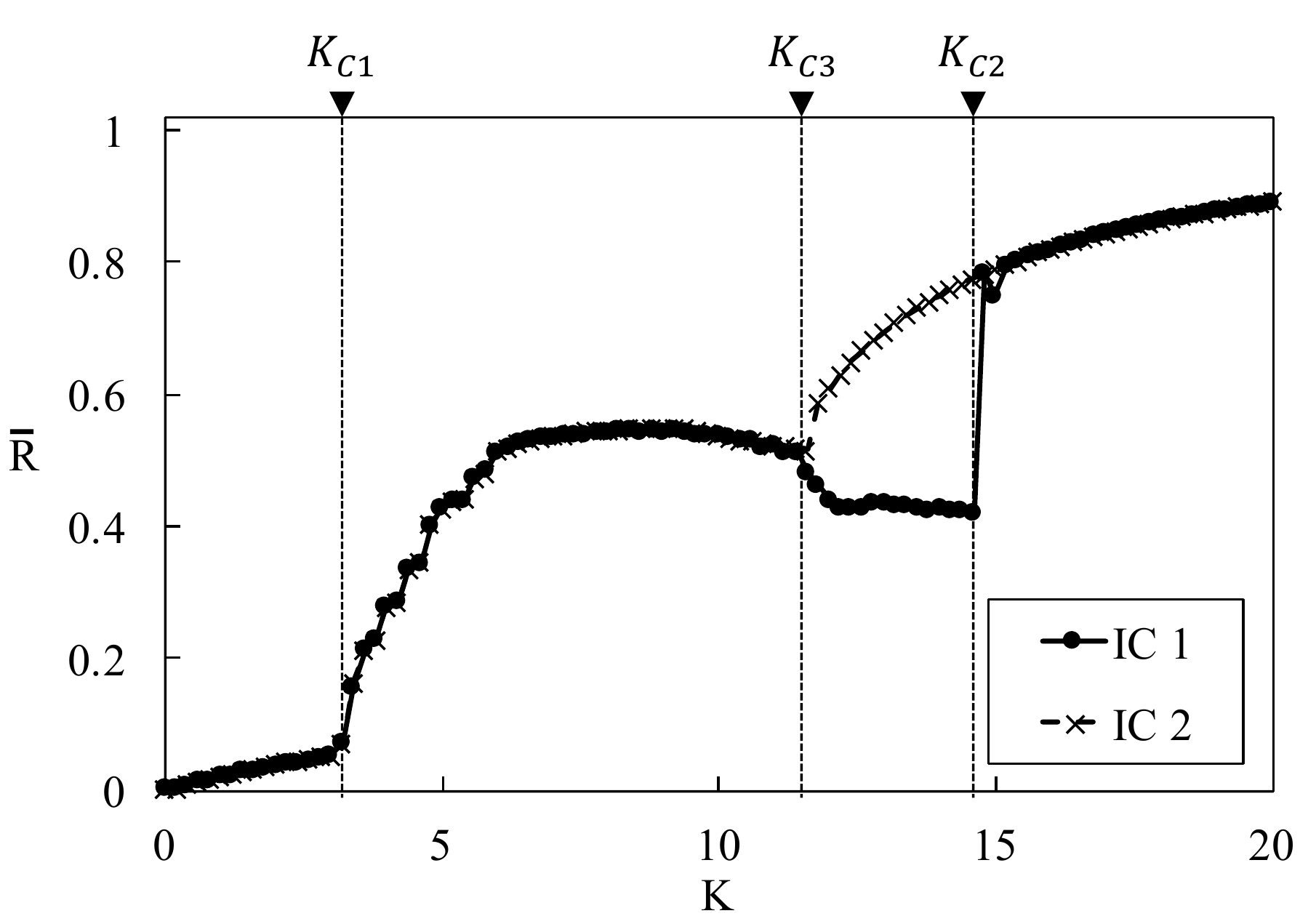}
            \caption{$\Delta = 0.6 \, (q\approx 0.42)$}
		    \label{}
        \end{subfigure}
            	\begin{subfigure}[c]{0.47\textwidth}
            \includegraphics[width=\textwidth]{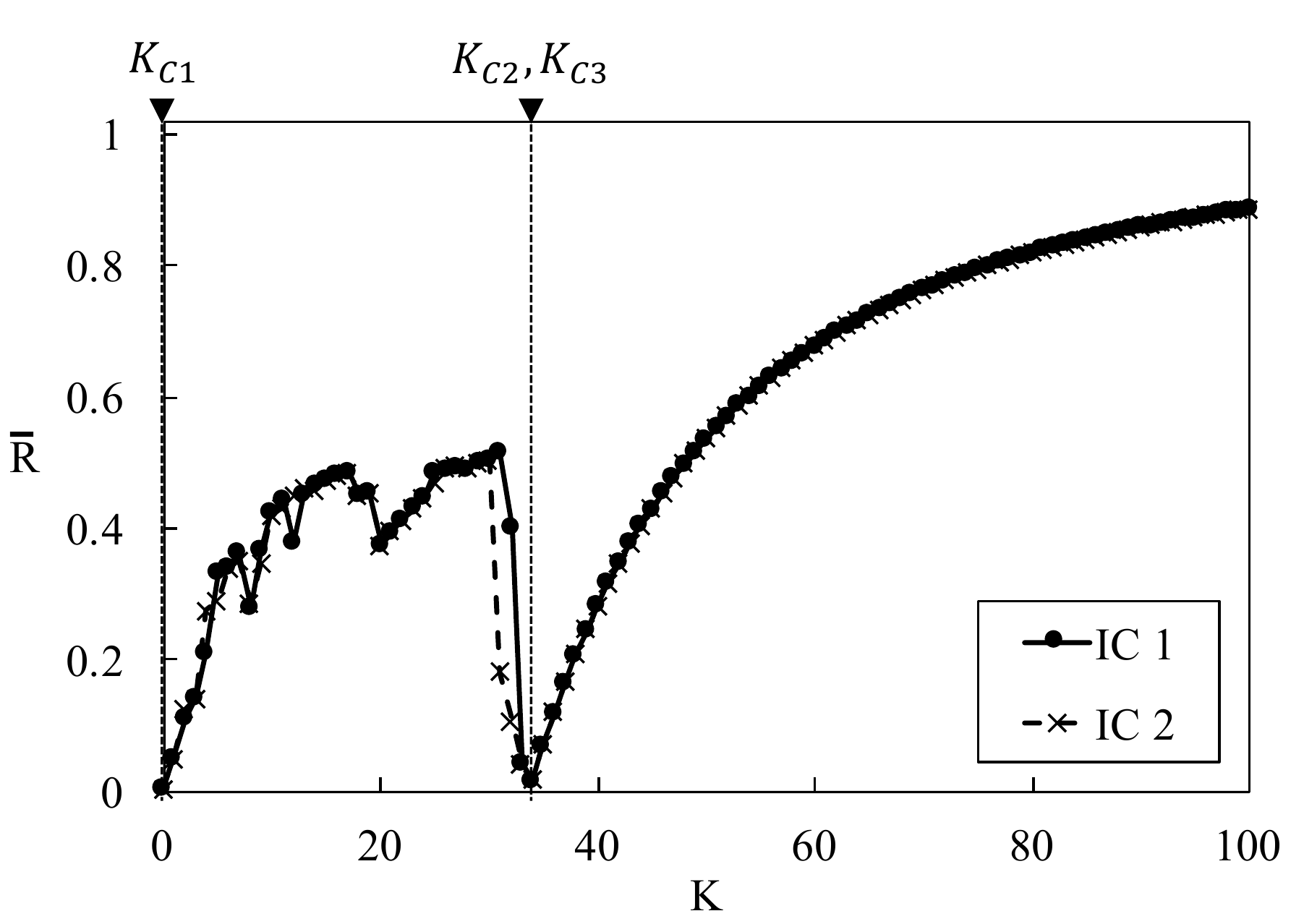}
            \caption{$\Delta = 0.2 \, (q\approx 0.14)$}
		    \label{}
        \end{subfigure}
        \quad
        \begin{subfigure}[c]{0.47\textwidth}
            \includegraphics[width=\textwidth]{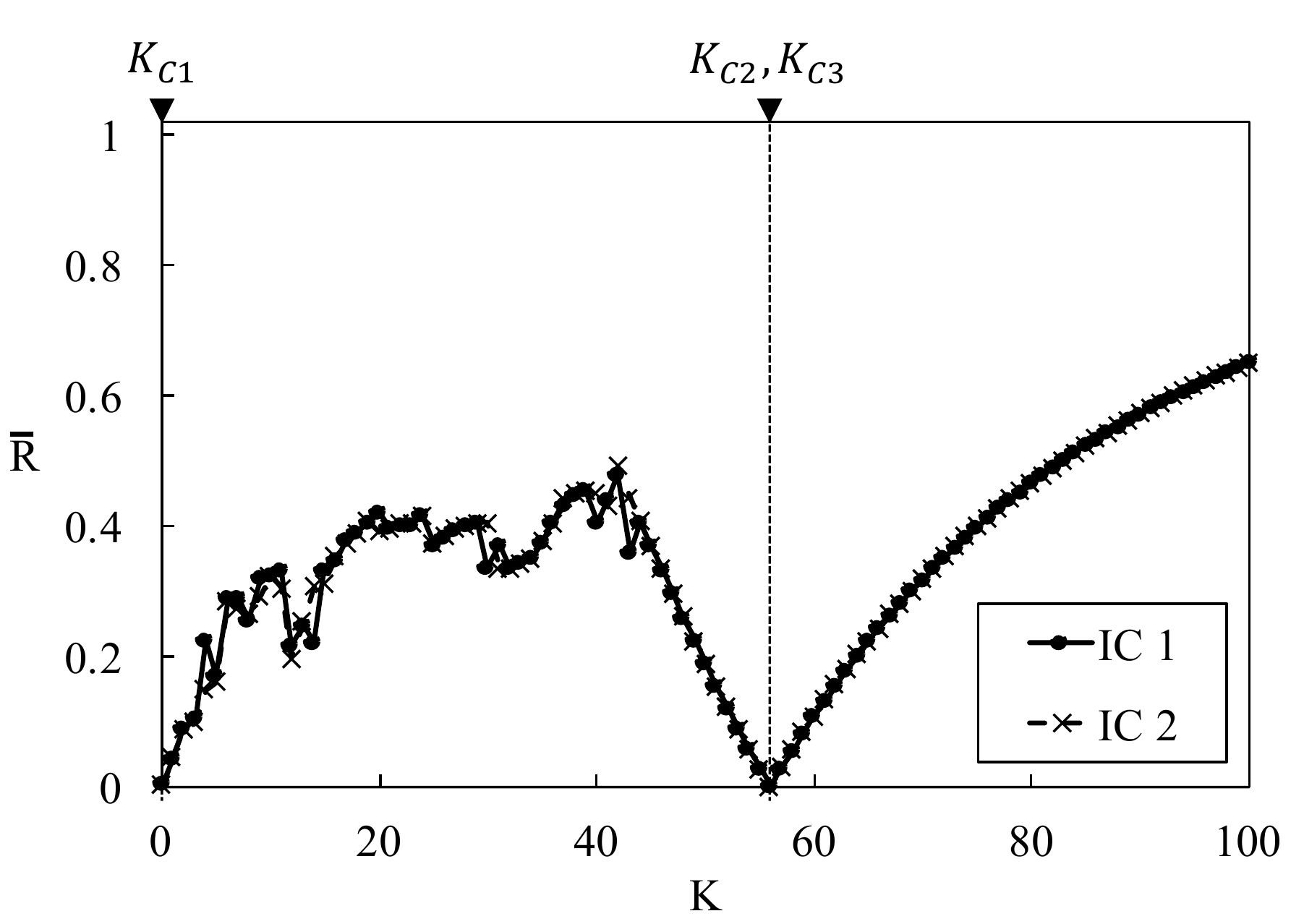}
            \caption{$\Delta = 0.2 \, (q\approx 0.14)$}
		    \label{}
        \end{subfigure}
    \caption{Comparison of the hysteretic loops of synchronization in case of different  kernel radii. Results with zeroth and first order kernel-functions are shown on the left and right, respectively.}
    \label{fig:hysteretic_loops}
\end{figure}

If $q$ is large enough (Figs. \ref{fig:hysteretic_loops}A-D), the spatially coupled model exhibits similar hysteretic behavior as the fully-coupled model. However, spatial coupling leads to the appearance of a \revised{transitional state} of the system (Fig. \ref{fig:hysteretic_loops}). Starting from an incoherent state (IC 1) and a small coupling parameter $K$, the average order parameter $\bar{R}$ stays near zero.  As $K$ is increased, at a critical point $K_{C1}$ the synchronization begins, i.e. $\bar{R}$ start to increase. Further increasing $K$ leads to $\bar{R}$ reaching a maximum and it stays constant for a range of $K$. However it decreases before transitioning to the globally coherent state by starting to increase again at a critical value of the coupling parameter $K_{C2}$. Therefore in case of IC 1, the globally coherent state is reached through two critical points $K_{C1}$ and $K_{C2}$, the beginning of the local and global synchronization, respectively. In case of IC 2, i.e. starting from a coherent state and decreasing the value of $K$, the system either jumps into an incoherent state (Fig. \ref{fig:hysteretic_loops}A) or the transitional state corresponding to locally locked clusters (Fig. \ref{fig:hysteretic_loops}B-D) at a bifurcation point $K_{C3}$. For $K<K_{C3}$ there is almost no difference between IC 1 and IC 2. In particular, if the maximal coupling strength is small enough, the local behavior dominates. 

Under a critical value of the relative kernel radius $q$, the local and global behavior are separated and the system exhibits no hysteresis (Figs. \ref{fig:hysteretic_loops}E-F), i.e. there is almost no difference between IC 1 and IC 2. Moreover, there is no constant part of the $\bar{R}-K$ diagram. Decreasing $q$ increases $K_{C2}$ and $K_{C3}$ and decreases $K_{C1}$. For $q\approx0.14$ (Figs. \ref{fig:hysteretic_loops}E-F) the synchronization starts near $K\approx 0$. 

 There is no qualitative difference between the results calculated with the applied kernel-functions. However, for constant $\Delta$, $W_{ij}^1$ leads to a higher level of dilution compared to $W_{ij}^0$. As a result, having the same kernel radii, the range of the partial synchronization (i.e. $K_{C1} \leq K \geq K_{C2}$) is always larger when using a first order than a zeroth order kernel-function. Interestingly, the average order parameter always decreased before the global synchronization began at $K_{C2}$ in all of our simulations. 
 
 \revised{The transitional state of the system is a result of the spatial coupling and the natural frequency distribution. Spatial coupling was also considered for 2D lattices by \cite{Odor2018} taking randomly distributed natural frequency values. While the system showed hysteresis there was no drop in the order parameter. Nevertheless, they observed no hysteresis for large system sizes, which is analogous with our results in Fig. \ref{fig:hysteretic_loops}e and f. Similar results was also reported in \cite{Tumash2018} for randomly diluted systems, showing that increasing dilution decreases the hysteretic region while increasing the mass leads to the opposite. In Fig. \ref{fig:hysteretic_loops} $K_{C2}-K_{C3}$ decreases for decreasing $q$ and eventually it reaches zero.}

\begin{figure}[H]
    \centering
        \begin{subfigure}[c]{\textwidth}
            \includegraphics[width=\textwidth]{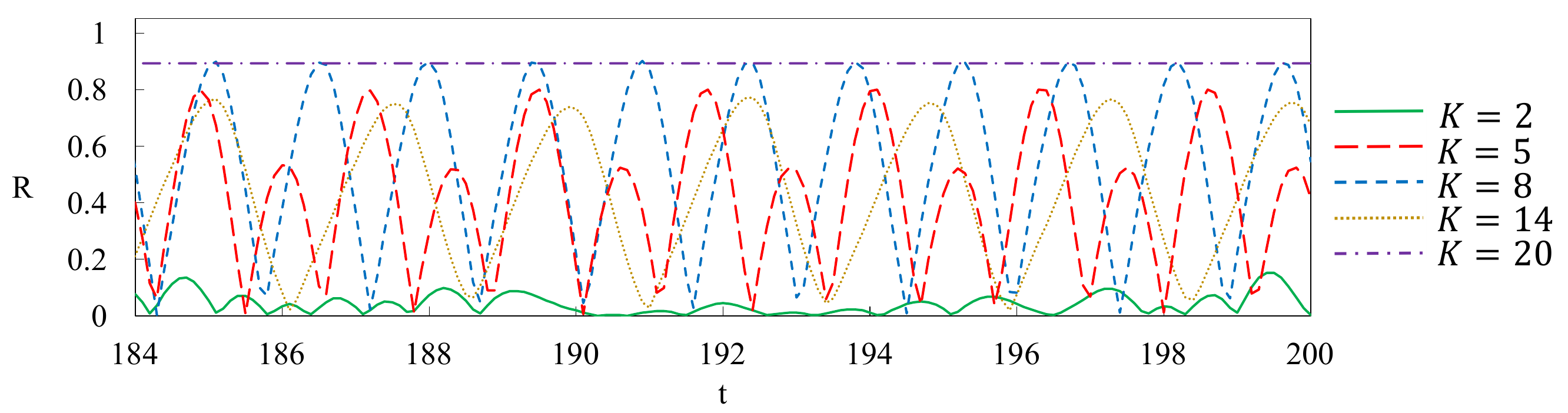}
            \caption{•}
		    \label{}
        \end{subfigure}
            \begin{subfigure}[c]{\textwidth}
            \includegraphics[width=\textwidth]{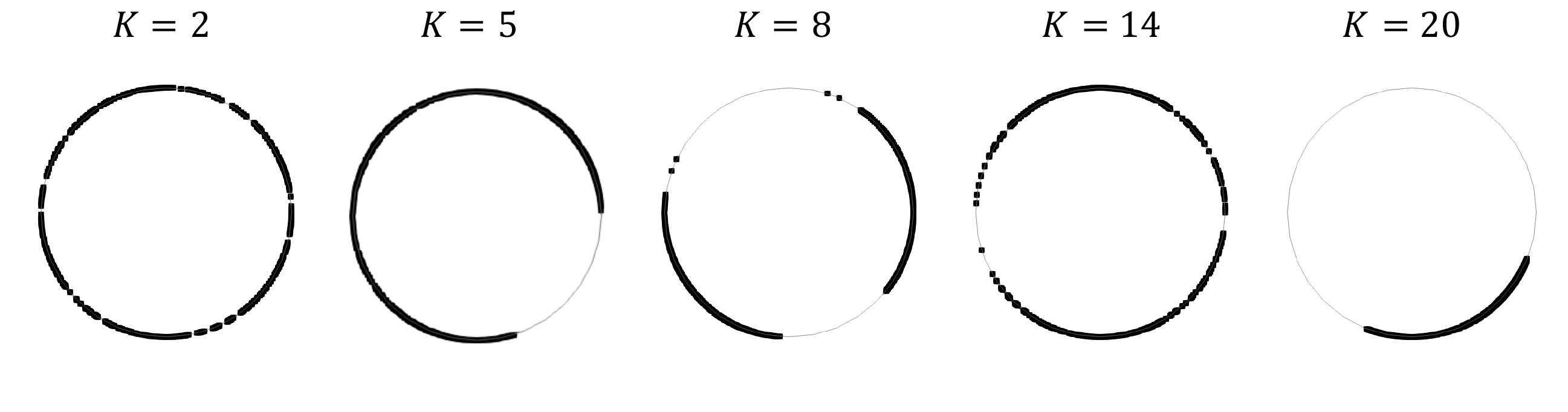}
            \caption{•}
		    \label{}
        \end{subfigure}
                \begin{subfigure}[c]{\textwidth}
            \includegraphics[width=\textwidth]{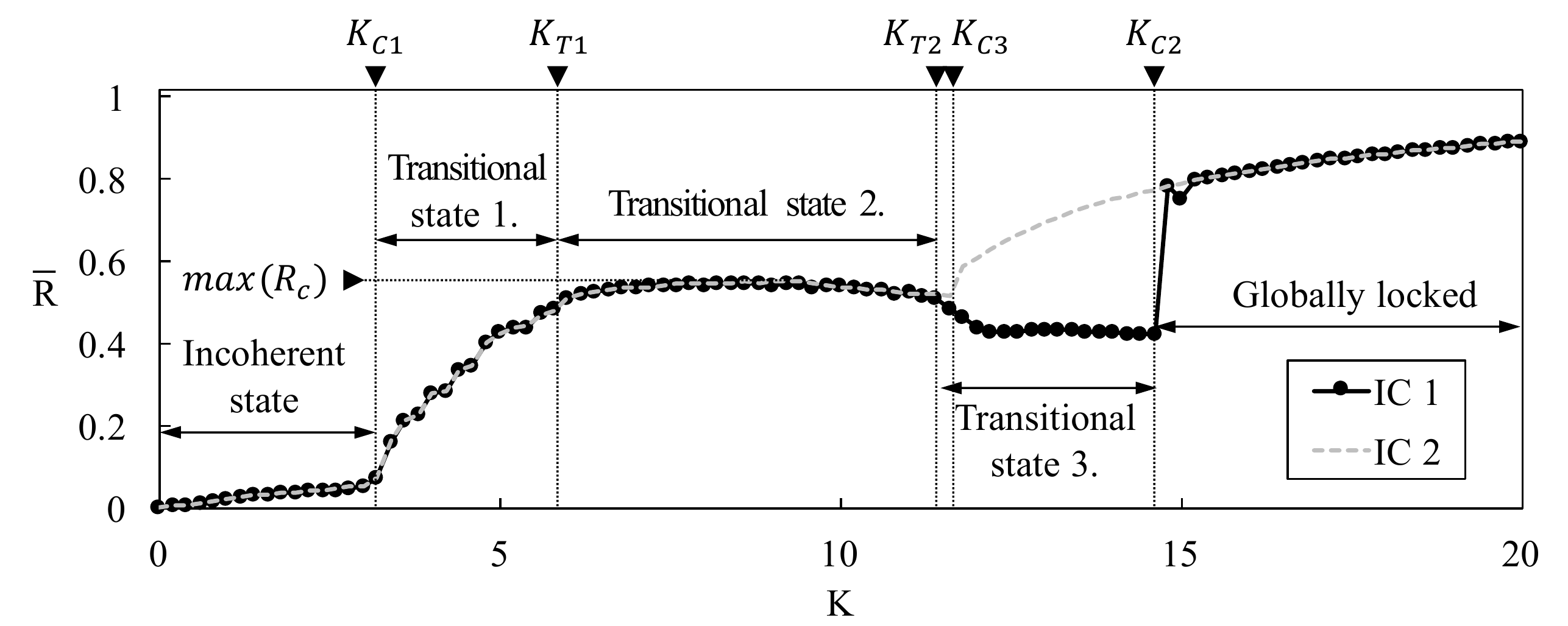}
            \caption{•}
		    \label{} 
        \end{subfigure}
    \caption{States of a system of spatially coupled oscillators depending on the coupling parameter $K$. The relative kernel radius is $q\approx0.46$, taking $\Delta = 0.6$ and $\Delta x = 0.05,N=400$ applying first order kernel-function. a) Time series of the order parameter $R$ for different values of $K$ calculated for IC 1. For $K_{C1} < K < K_{C2}$, $R$ has an oscillatory nature. b) Final phases of the oscillators in polar form at $t=500s$ illustrating the different states of the system calculated for IC 1. c) States of the system and the time average $\bar{R}$ of the order parameter depending on the coupling parameter $K$.}
    \label{fig:states}
\end{figure}

\begin{figure}
    \centering
    \includegraphics[width=0.8\textwidth]{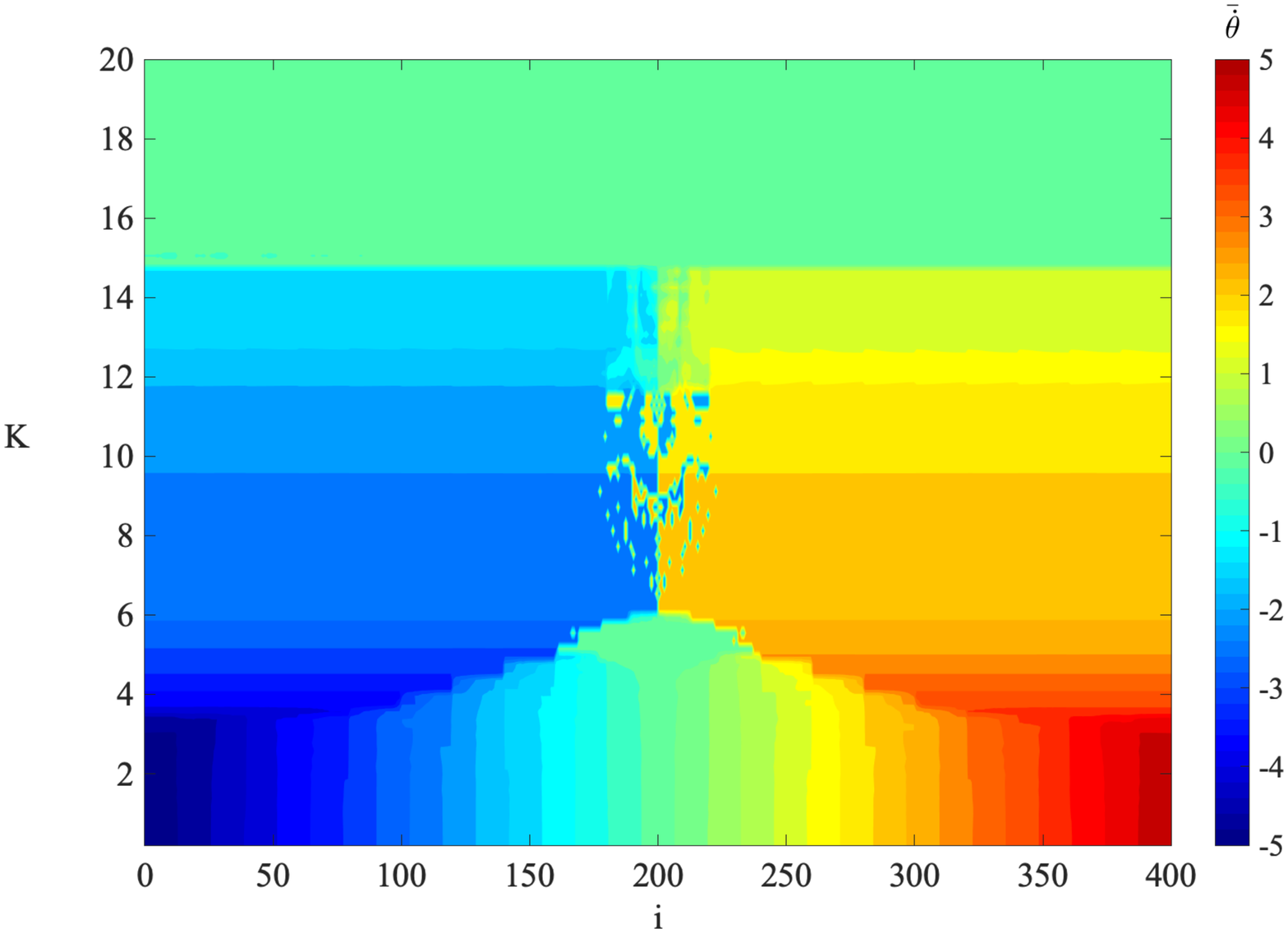}
    \caption{\revised{Average frequency $\bar{\dot\theta}$ in the function of the oscillator index $i$ and the coupling strength $K$.}}
    \label{fig:avg_frequency}
\end{figure}

\begin{figure}
    \centering
    \includegraphics[width=0.6\textwidth]{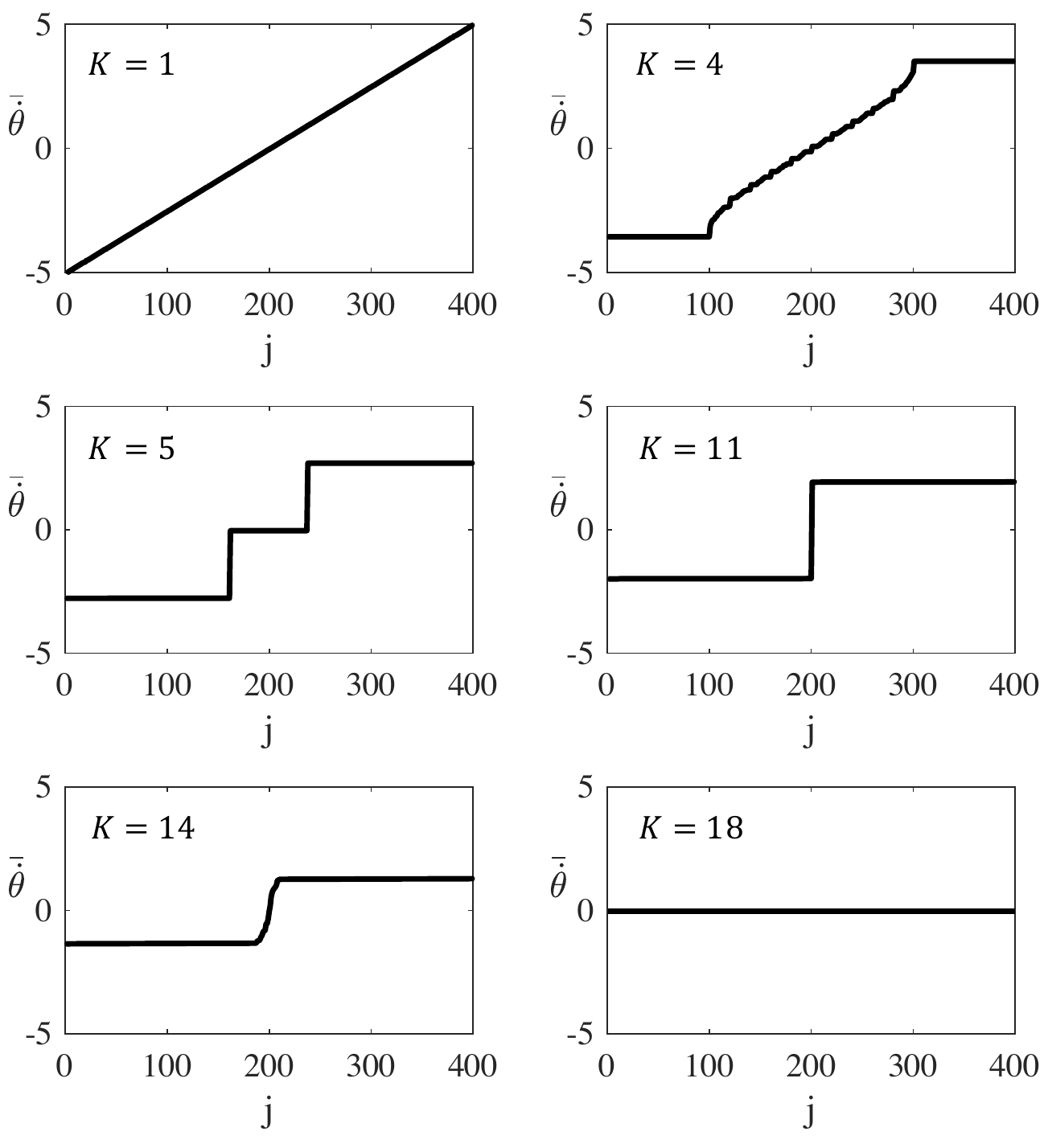}
    \caption{\revised{Average frequency $\bar{\dot\theta}$ of the oscillators for different $K$ values, ordered for increasing frequency values. $j$ is the oscillator index after reordering for each $K$ values. For small $K$ values, the system is in an unsynchronized state ($K=1$), then for increasing $K$ locally locked clusters develop and merge ($K=4$ and $K=5$) until there are only two separated clusters ($K=11$) which start to merge ($K=14$) and finally there is only one cluster ($K=18$).}}
    \label{fig:reordered_frequency}
\end{figure}

\subsection{States of the system}
We analyzed the states of the spatially coupled model depending on the coupling parameter $K$ in case $q$ is large enough for the system to exhibit the hysteretic behavior. Analysis of the solutions of Eq. \ref{eq:inertia_kuramoto_kernel} in time was carried out by taking uniformly diffused initial conditions (IC 1) and first order kernel-function $W_{ij}^1$ with fixed kernel radius $\Delta=0.6$ ($q\approx 0.46$) and $\Delta x = 0.05$. By examining the time evolution of the order parameter $R$ (Fig. \ref{fig:states}A) and the phases of the individual oscillators (Fig. \ref{fig:states}B), three states of the system in the $\bar{R}-K$ diagram (Fig. \ref{fig:states}C) was found. \ref{fig:states}B is only a visualization of the data and the distances measured on the circle are independent of the strength of the interactions. \revised{Fig. \ref{fig:avg_frequency} illustrates the different states by showing the average frequency in the function of the oscillator index and the coupling strength. Fig. \ref{fig:reordered_frequency} shows the average frequency in the function of the oscillator index reordered for increasing frequency values for $K$ values corresponding to the different states. }\revised{Starting from the \emph{incoherent} state, the system goes through \emph{transitional} states with locally locked oscillators until it reaches a \emph{globally locked state}. The transitional state is further divided into three parts characterized by different frequency distribution patterns (Fig. \ref{fig:reordered_frequency})}.
\revised{Note, that the states of the system strongly depend on the natural frequency distribution (Eq. \ref{eq:omega}) which in this case induces a correlation between the index $i$ and $\dot{\theta}_i$.}


If $0 < K < K_{C1}$, the system is in the incoherent state. Since we keep the natural frequency distribution defined by Eq. (\ref{eq:omega}) and it is a dominant part of the equation, there are two \revised{sub-populations} in the initial configuration ($\Omega_i < 0$ and $\Omega_i\geq0$). The order parameter $R$ varies in time in a random-like manner, but it remains close to zero (Fig. \ref{fig:states}A, $K=2$) and the phases of the oscillators are heterogeneous (Fig. \ref{fig:states}B $K=2$). The fluctuations of $R$ in time can be attributed to the finite number of oscillators in the system. \revised{The average frequency of the oscillators is different for each oscillators (Fig. \ref{fig:avg_frequency}).} Increasing $K$, the average order parameter $\bar{R}$ stays near zero until reaching a $K_{C1}$ value. Above $K_{C1}$, the degree of synchronization gradually increases.

\revised{If $K_{C1} < K < K_{C2}$ locally locked clusters develop and the system start to transition to the globally locked state.} Increasing $K$ from $K_{C1}$, \revised{ the locally locked clusters grow and merge. This process continues, until two well separable locally locked clusters develop at a $K_{T1}$ value. This state of the system corresponds to a travelling wave. Between $K_{T1}$ and $K_{T2}$, the average order parameter reaches a maximum $max(R_C)$ and it stays constant while $R$ oscillates in time (Fig. \ref{fig:states}A, $K=8$). This state is a standing wave and it remains until $K_{C2}$.} Note, that the overlap of the clusters neither result in their interaction nor global synchronization. \revised{Above a $K_{T2} > K_{T1}$ value, the clusters start to affect each other leading to their merge (Fig. \ref{fig:states}B, $K=14$) but the system still corresponds to a standing wave. While $R$ is still oscillating in time (Fig. \ref{fig:states}A, $K=14$), the merging of the local clusters is indicated by a decrease in the maximal value of the order parameter.} Further increasing $K$, the system rapidly jumps into a globally locked state in a bifurcation point, at a $K_{C2}$ value. At this point, the clusters merge as one of them take over the domination \revised{and the average frequency becomes zero (Fig. \ref{fig:avg_frequency}).}


If $K_{C2} < K$, there is only one cluster, the system is in a globally locked state (Fig. \ref{fig:states}B, $K=20$) and $R$ is constant in time (Fig. \ref{fig:states}A, $K=20$). As $K$ is further increased, the synchronization and $\bar{R}$ gradually increases (Fig. \ref{fig:states}A-B, $K=200$) and the system seemingly reaches the globally coherent state at a large value of $K$. In contrast to the fully coupled model, where $R\approx1$ is reached at moderate $K$, here the results imply, that $\bar{R}\rightarrow1$ asymptotically. Accordingly, spatial coupling can hinder full synchronization for large system sizes in case the coupling strength of the individual oscillators is limited. 

We also checked the solutions of the model for IC 2. For $0 \leq K \geq K_{C3}$, the system has the same states as for IC 1. Above $K_{C3}$, there is one globally locked cluster with constant $R$ in time. 

\begin{figure}
    \centering
    \includegraphics[width=0.7\textwidth]{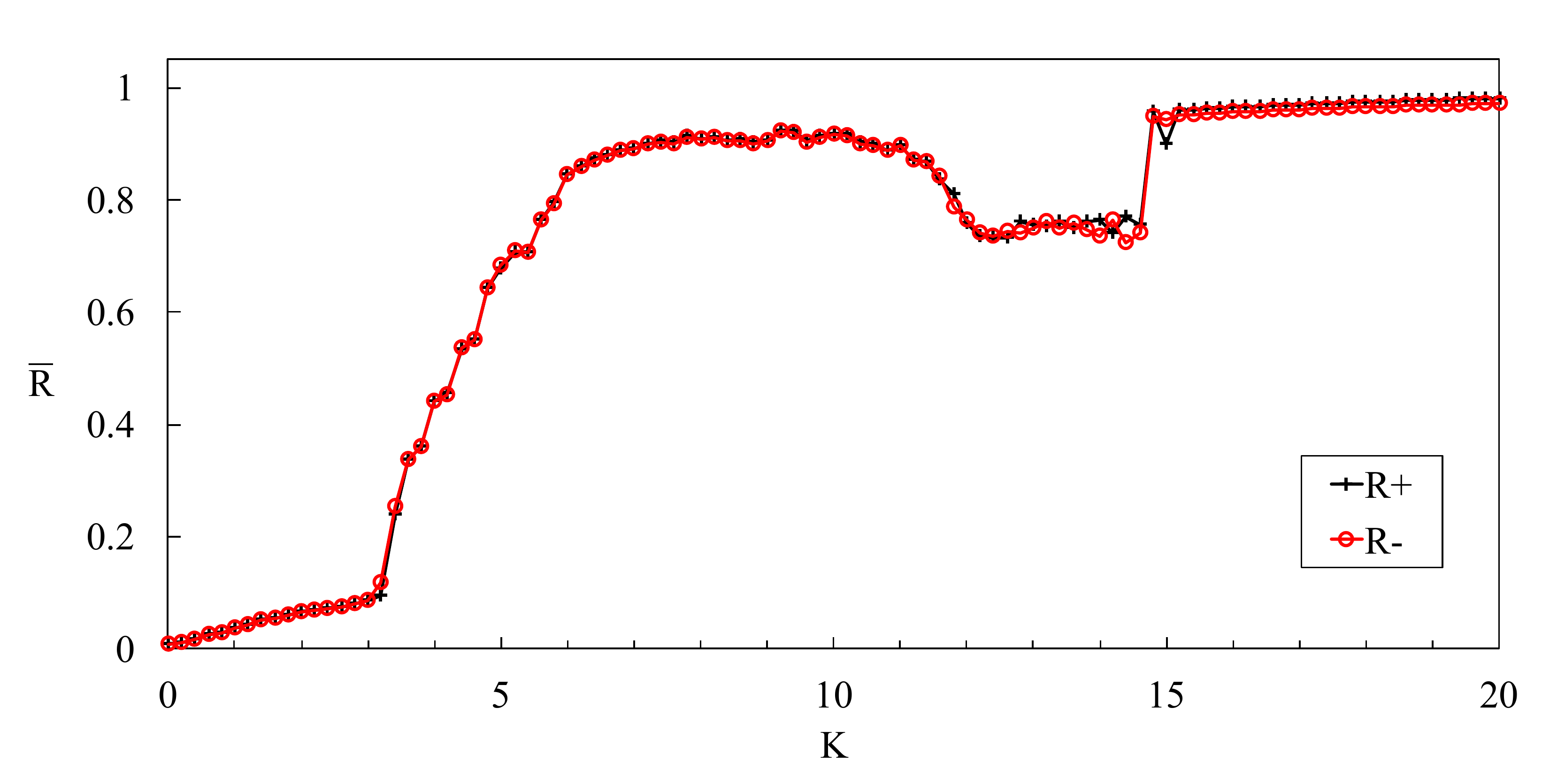}
    \caption{\revised{Average order parameter in the function of the coupling strength for the sub-populations corresponding to $\Omega_i < 0$ (R-) and $\Omega_i \geq 0$ (R+) for IC 1 taking the same parameters as in Fig. \ref{fig:states}c.}}
    \label{fig:rpm}
\end{figure}

\begin{figure}
    \centering
    \includegraphics[width=0.7\textwidth]{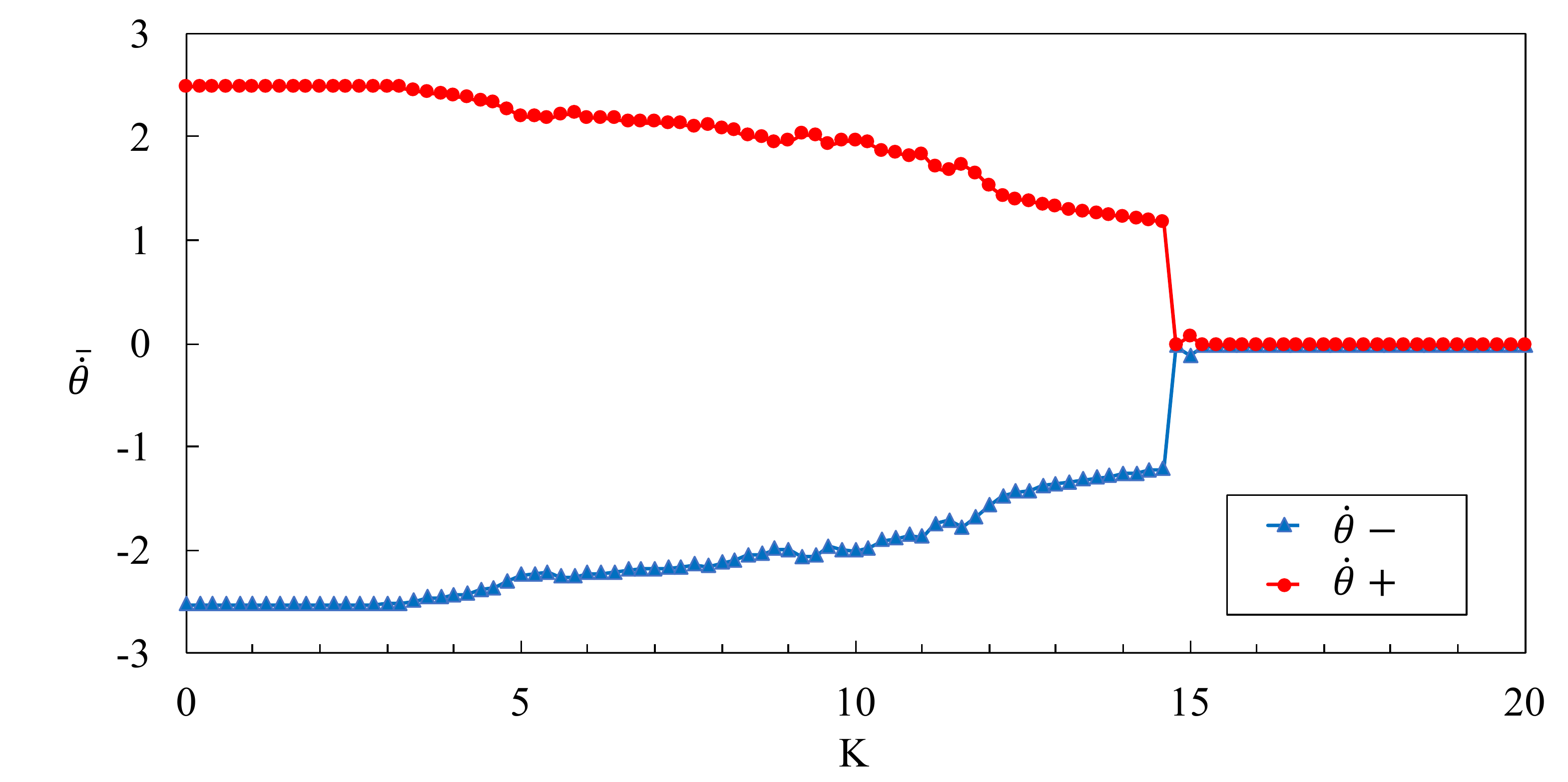}
    \caption{\revised{Average frequency in the function of the coupling strength for the sub-populations corresponding to $\Omega_i < 0$ ($\bar{\dot\theta}_i$-) and $\Omega_i \geq 0$ ($\bar{\dot\theta}_i$+) for IC 1 taking the same parameters as in Fig. \ref{fig:states}c.}}
    \label{fig:thetapm}
\end{figure}

\revised{Due to the initial condition regarding the natural frequency distribution, it is worth to handle the sub-populations corresponding to $\Omega_i < 0$ and $\Omega_i \geq 0$ separately. Figs. \ref{fig:rpm} and \ref{fig:thetapm} illustrate the order parameter and the average frequency for the different sub-populations. According to the diagrams, the two sub-populations behave symmetric as K is increased.} 

\subsection{Effect of the relative kernel radius}
In Fig. \ref{fig:hysteretic_loops}, $\Delta$ and $q$ were varied simultaneously in the different simulations. The diagrams show varying values of $K_{C1},K_{C2},K_{C3}$ and different $max(R_C)$ in the locally locked state. To examine the effect of the parameters separately, we calculated $\bar{R}$ versus $K$ for both IC 1 and IC 2 keeping either the number of oscillators in a neighborhood of a general point constant (i.e. keeping $\Delta$ and $\Delta x$ constant) or the relative kernel radius $q$ fixed. Since, there is no qualitative difference between the first and zeroth order kernel-functions, we applied $W_{ij}^1$ in all cases. 
\begin{figure}[h!]
    \centering
    \includegraphics[width=0.85\textwidth]{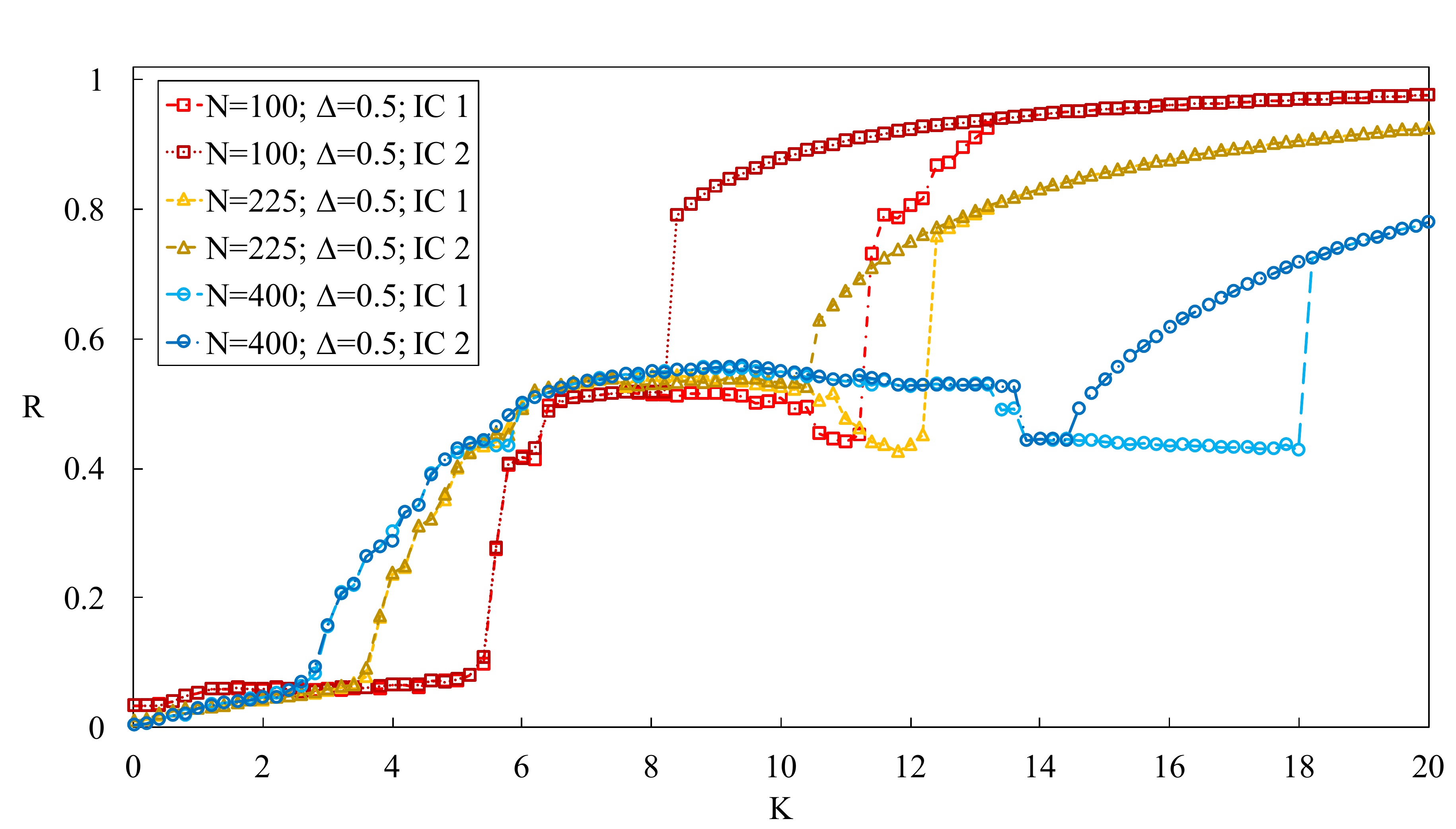}
    \caption{Average order parameter $\bar{R}$ versus the coupling parameter $K$ calculated for IC 1 and IC 2 fixing the kernel radius $\Delta = 0.5$ and varying the domain size $L_1=0.5, L_2=0.75, L_3=1$ ($q_1\approx0.7,q_2\approx0.47,q_3\approx0.35$), applying the first order kernel $W_{ij}^1$.}
    \label{fig:fix_delta}
\end{figure}
\begin{figure}[h!]
    \centering
    \includegraphics[width=0.85\textwidth]{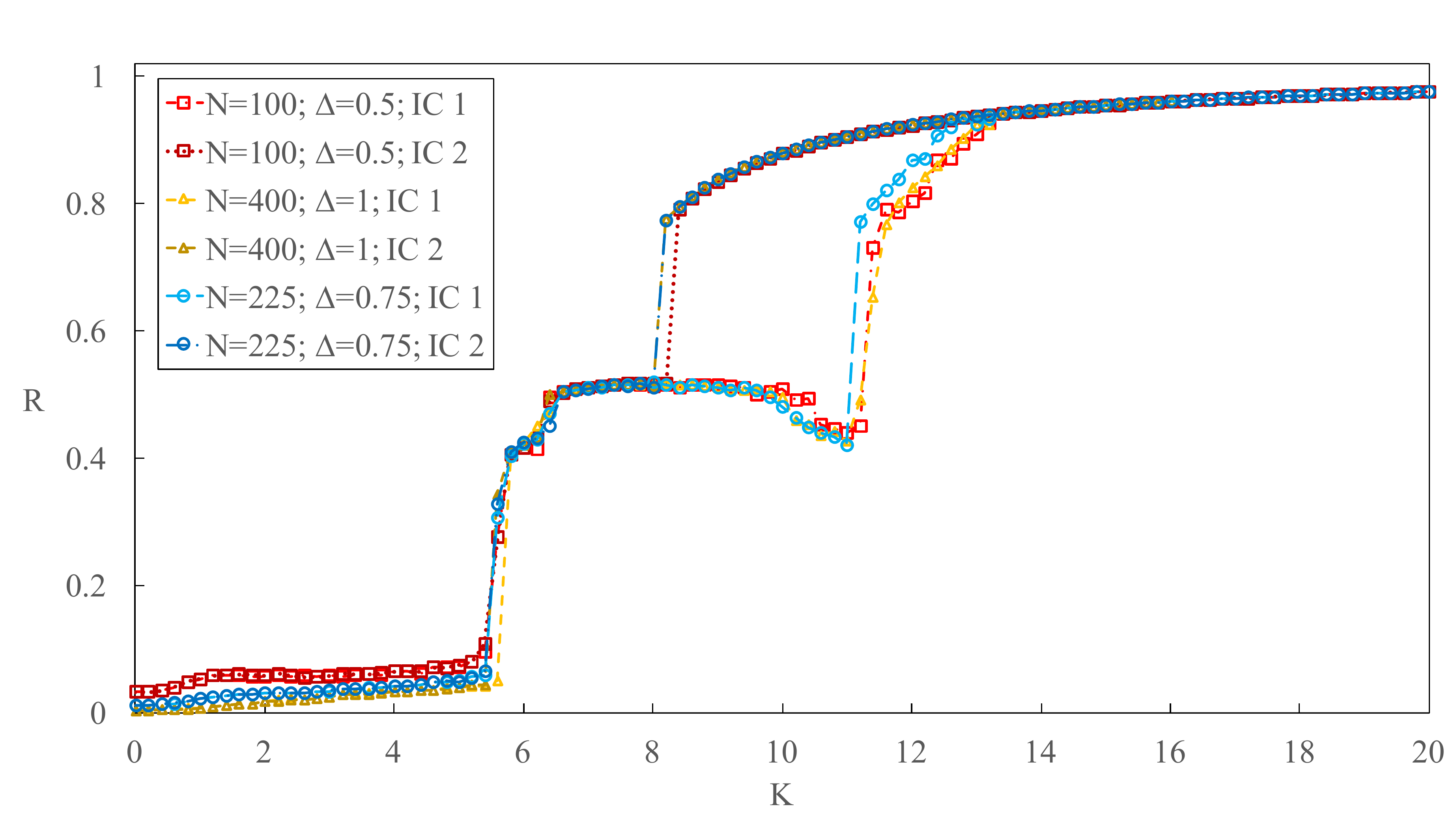}
    \caption{\revised{Average order parameter $\bar{R}$ versus the coupling parameter $K$ calculated for for IC 1 and IC 2, fixing $q\approx0.7$, and changing both the kernel radius and the domain size: $\Delta_1 = L_1 = 0.5, \Delta_2 = L_2 = 0.75, \Delta_3 = L_3 = 1$, applying the first order kernel $W_{ij}^1$.}}
    \label{fig:fix_q}
\end{figure}
\revised{First, we considered three systems of oscillators with different domain sizes, but the same number of particles in a neighborhood of a general point of the domain. The domain sizes and the particle numbers were chosen to be $L_1=0.5, N_1=100$, $L_2=0.75, N_2=225$ and $L_3=1, N_3=400$. The grid cell size and the kernel radius were $\Delta x = 0.05$ and $\Delta=0.5$ in all systems. Accordingly, the relative kernel radii were $q_1\approx0.7, q_2 \approx 0.47, q_3 \approx 0.35$.} As we can see in Fig. \ref{fig:fix_delta}, although the number of oscillators in a general in-domain neighborhood is the same in all cases, both the local and global parts of the diagram are affected. In case of a smaller relative kernel radius, the local synchronization starts at smaller values of $K$ and a larger $K$ is necessary for the onset of the global synchronization. Furthermore, the maximal order $max(R_C)$ of the system in the locally synchronized state is higher for $q_2$. Smaller relative kernel radius $q$ results in a larger range of $K$ corresponding to the locally synchronized states (i.e. $|K_{C1}-K_{C2}|$ increases), but it also requires a larger value of $K_{C3}$ for the global synchronization.

\revised{Next, we fixed $q$ and $\Delta x = 0.05$ by choosing $L_1 = \Delta_1 = 0.5, N_1 = 100$, $L_2 = \Delta_2 = 0.75$, $L_3 = \Delta_3 = 1, N_3 = 400$ ($q_1 = q_2 = q_3 = \approx 0.7$).} Fig. \ref{fig:fix_q} shows, that there is only a small difference between the locations of $K_{C1}$ and $K_{C3}$, but the transitional state is not affected. The onset of the local synchronization requires larger $K$ as the system size is increased, but the global synchronization can be maintained for smaller values of $K$. Similar dependencies on the system size were reported in \cite{Tanaka1997,Olmi2014} for the fully coupled and diluted systems, respectively. These results suggest, that the $\bar{R}-K$ diagram is characterized by $q$ and the local behavior is not affected by the system size. The maximal order $max({R_C})$ of the system in the locally synchronized state was also independent of the system size in our simulations. 

\begin{figure}[h!]
    \centering
    \includegraphics[width=\textwidth]{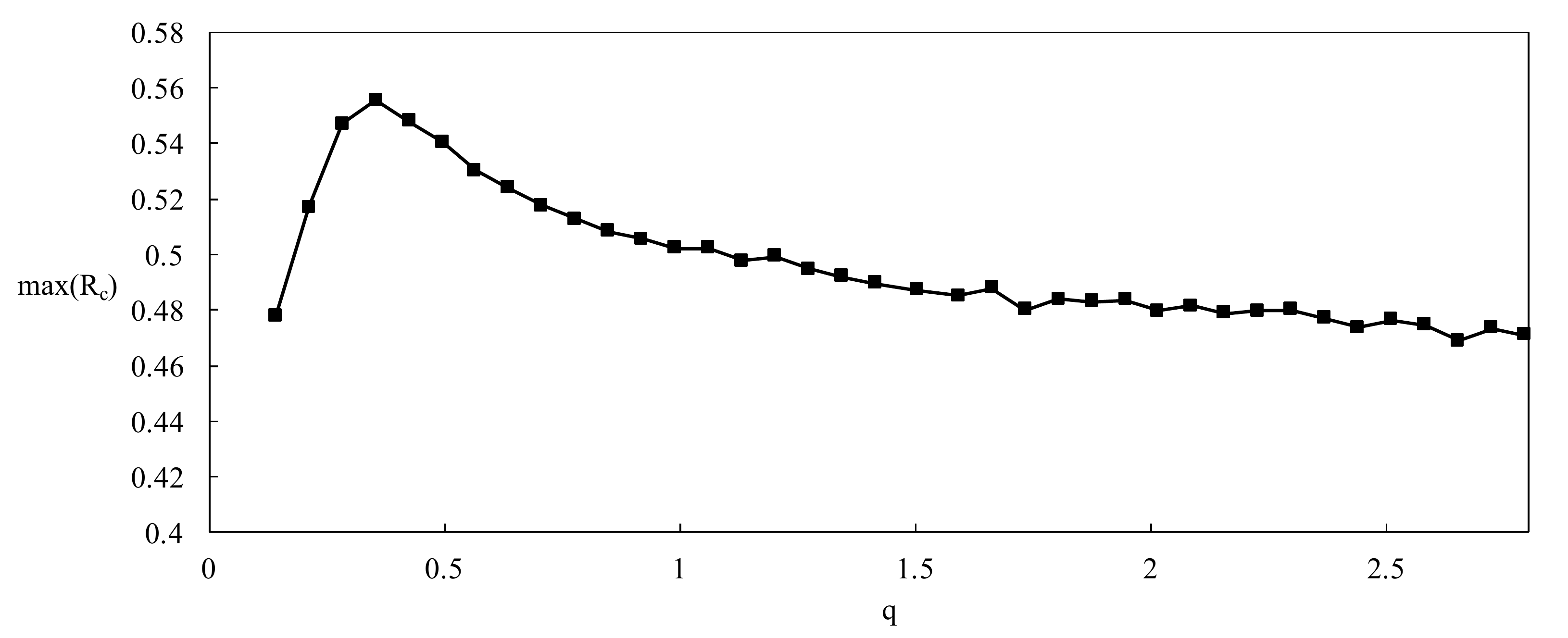}
    \caption{Maximum of the average order parameter in the locally locked state $max(R_C)$ depending on $q$ fixing $L=1, \Delta x = 0.05, N=400$, applying first order kernel-function. For the simplicity, the diagram was computed assuming IC 1.}
    \label{fig:rmax_delta}
\end{figure}

Finally, we computed the maximal value of the order parameter in the locally locked state $\mathrm{max}(R_C)$ for $K_{C1} < K < K_{C2}$ depending on $q$, keeping $L=1, \Delta x = 0.05, N=400$ fixed and applying first order kernel-function. According to Fig. \ref{fig:rmax_delta}, the $max(R_C)(q)$ function has a maximum. Consequently, there is an optimal value of the relative kernel radius in terms of the local synchronization. If $q$ is too large, the local behavior dominates and if the kernel radius is too small compared to the domain size, the local synchronization is hindered.

\section{Conclusion}
In summary, we carried out a numerical study to examine the effect of the spatial coupling of phase oscillators in the Kuramoto model with inertia. A finite size system was considered with compact zeroth and first order kernel functions to incorporate distance-dependent coupling strength between the oscillators. A dimensionless parameter, the relative kernel radius $q$ was introduced, expressing the ratio between the kernel radius and the maximal distance between the $i$th and $j$th oscillators. We examined the order of the global synchronization depending on the coupling parameter for different $q$ values. In the case of large kernel radii, the model gives back the original, fully coupled model.

We implemented the model in the Nauticle general purpose particle-based simulation tool \cite{Toth2017} and examined the hysteretic behavior of the spatially coupled model. We fixed the domain size and investigated the behavior for different values of $q$. For a range of $q$, the system exhibits a hysteretic behavior depending on the initial conditions and the coupling parameter, but for a range of $K$ values locally locked clusters develop and dominate. In case of the presence of the local clusters, the order parameter $R$ is oscillates in time.  We pointed out, that this oscillations can be attributed to locally locked clusters that develop and break up through some transitional states. If $q$ is small enough, the system exhibits no hysteresis, which agree with previous works showing that reducing the number of links between the oscillators lead to the vanish of the hysteretic behavior \cite{Olmi2014}. Analysis of the parameters in the problem showed, that the local behavior is characterized mainly by the relative kernel radius $q$, which has an optimal value leading to a maximal value of the synchronization in the locally locked state.

Our results suggest, that large systems of oscillators having a distance limited vision, such as the fireflies can develop locally synchronized clusters without transitioning into global synchronization. The proposed framework can be used to model more complex spatially coupled systems. Some bacteria such as \emph{Myxococcus xanthus} \cite{Leonardy2007} or systems of microgears \cite{Aubret2018} exhibit not only social interactions but also complex mechanical behavior.  By the implementation of the model into Nauticle it is possible to consider a mathematical coupling of the Kuramoto model with other equations, such as mechanical models to examine complex systems of spatially moving oscillators.

\section*{Acknowledgement}

The research reported in this paper was supported by the Higher Education Excellence Program of the Ministry of Human Capacities in the frame of Water Science \& Disaster Prevention research area of Budapest University of Technology and Economics (BME FIKP-V\'{I}Z). The research reported in this paper has been supported by the National Research, Development and Innovation Fund (TUDFO/51757/2019-ITM, Thematic Excellence Program).
The research reported in this paper has been supported by the National Research, Development and Innovation Fund (TUDFO/51757/2019-ITM, Thematic Excellence Program).

\bibliographystyle{unsrt}
\bibliography{references}
\end{document}